# Radial Variation of Microstructure in a Direct-Chill Cast AA7050 Billet on Homogenization

Pikee Priya[a,#], Kyle Fezi[a], D. R. Johnson[a], M. J. M. Krane[a]

[a] Purdue Center of Metal Casting Research, School of Materials Engineering, Purdue University, West Lafayette, IN, USA.

[#]ppriya@iisc.ac.in

**Abstract**

A "through-process" solidification homogenization numerical model to study the microstructural evolution has been developed for AA7050 alloy. A continuum scale Direct Chill Casting (DCC) solidification model has been coupled with a meso-scale homogenization precipitation model to evaluate the radial variation of microstructure in a cylindrical DC cast billet after homogenization. The Local Solidification Times predicted by the DCC numerical model is used to predict the Secondary Dendritic Arm Spacing (SDAS) across the radius from an empirical relationship. It also predicts the macrosegregation used to estimate the radial as-cast microstructures (phases and phase fractions) using Thermo-Calc$^{TM}$. Macrosegregation affects the initial and homogenized microstructures across the radius, making the surface prone to recrystallization, due to lesser dispersoids and larger precipitated particles during cooling, caused by lesser extent of Zenner pinning and Particle Stimulated Nucleation (PSN) of recrystallized grains respectively.

## 1. Introduction

A comprehensive study of heat treatable aluminum extrusions starts at casting of the metal and ends at evaluating the performance of aluminum extrusions after heat treatment and thermo-mechanical treatment. The various processes studied are casting, homogenization, and thermo-mechanical treatment. Each process affects the ones downstream, making the study complex and involving a wide range of length and time scales. There is a complex interplay of processing, microstructure and properties which must be understood. This is what forms the basis for "Integrated Computational Materials Engineering" (ICME).

"Through process modeling" is not a new concept and has been increasingly used in different fields. In the field of metal processing, several researchers have attempted it in the past. Solidification and homogenization are two very closely related processes and have been studied by many researchers. Early studies by Brooks et al.[1] evaluated the evolution of the microsegregation during casting and homogenization of stainless steel welds. It essentially involved solving the mass diffusion equations in a cylindrical domain with a given temperature history with movement of the solid-liquid interface boundary triggered by phase equilibria. The DICTRA$^{TM}$ software, which can predict one dimensional diffusion induced phase transformations, has increasingly been used to predict solidification and homogenization microstructures in various alloy systems. Lippard et al.[2] and Samaras and Haidemenopoulos[3] used DICTRA$^{TM}$ to predict microsegregation and phase fraction evolutions during casting and homogenization of AerMet100 steel and AA6061, respectively. The Pseudo-Front-Tracking method[4] was used by Gandin and Jacot[5] to model solidification and homogenization in AA3003 alloy which was coupled with a precipitation model to predict width of the precipitate-free zones. Warnken et al.[6] used phase-field methodology to study evolution of as-cast microstructure and homogenization in nickel-based alloys.

Some of the larger length-scale through-process modelling studies include works by Neumann et al.[7] and Tin et al.[8] who modeled processing of aluminum sheets and Ni-based superalloy discs, respectively. Neumann et al.[7] modelled casting, homogenization and forming of the Al sheets, where the model in each step produced results which, along with some experimental results, were fed to the next model to create a through process model. For instance, the casting model predicted grain size and microsegregation which, along with the experimentally measured grain size distribution, was fed into the homogenization model. Tin et al.[8] described an integrated model to predict grain structure and defects during various processing stages of a gas turbine disc of INCONEL alloy 718. The process-stream that was studied were Vacuum Arc Remelting (VAR), homogenization, cogging, forging, and heat treatment.

The current study combines the numerical study of the first two processing stages of aluminum extrusions, namely casting and homogenization. These processes are studied for a DC-cast cylindrical billet of AA7050. The DC-casting solidification model in the continuum scale developed by Fezi et al.[9] feeds the radial microstructural and composition information to the meso-scale homogenization and precipitation models developed in work by Priya et al.[10,11] to determine the radial variation of microstructure after homogenization and cooling of the billet under industrial conditions. While the solidification model directly predicts the macrosegregation, giving the radial variation of compositions in the billet, the Secondary Dendrite Arms Spacing (SDAS) is indirectly estimated based on the predicted solidification times. This microstructural information helps us predict the microsegregation and volume fraction of the interdendritic phases in the as-cast microstructure. from Thermo-Calc$^{TM}$. This microsotructure then serves as the initial condition for the homogenization and post-homogenization cooling microstructure evolution prediction.

The radial variation of the as-cast microstructure in a cylindrical billet causes a variation in microstructural evolution during homogenization. A radial variation of homogenization temperature history during industrial processing conditions is also considered. A homogenization schedule applicable for the entire cross section of the billet, without causing remelting of the secondary phases has been proposed. The meso-scale homogenization model with experimental validation along with its extension to "through-process" modeling of DCC billets can be found in detail in dissertation work by P. Priya[13].

## 2. Methods

### 2.1. Domain Description

The heat transfer in a cylindrical billet under industrial homogenization conditions is modeled. The domain is axisymmetric with radius of 0.35 m and the boundary conditions shown in Figure 2. The conduction of heat in the billet is modeled through the heat conduction equation in cylindrical coordinates as:

$$\frac{1}{\alpha}\frac{\partial T}{\partial t} = \frac{1}{r}\frac{\partial}{\partial r}\left(r\frac{\partial T}{\partial r}\right) \tag{1}$$

where $\alpha = k_c/\rho c_p$, $k_c$ being the thermal conductivity, $\rho$ the density and $c_p$ the specific heat of AA7050 alloy. A symmetry condition is applied at the centerline.

$$\frac{\partial T}{\partial r} = 0 \; at \; r = 0 \tag{2}$$

The convective and radiation heat losses at the boundary are considered at billet surface as

$$-\frac{\partial T}{\partial r} = \frac{h}{k_c}(T_{R_0} - T_\infty) + \varepsilon\sigma(T_{R_0}^4 - T_\infty^4) \; at \; r = R_0 \tag{3}$$

$h$, being the convective heat transfer coefficient, $T_{R_0}$, temperature at outer radius, $R_0$ of the billet, $T_\infty$, the atmospheric temperature, $\varepsilon$, emissivity of the billet and $\sigma$, the Stefan Constant.

Table 1: Values of the heat transfer parameters used.

| Parameters | Values[12] |
|---|---|
| $\alpha$ | $6.24 \times 10^{-5}$ m²/s |
| $k_c$ | 153 W/Mk |
| $h$ | 10 W/m²K (heating) |
| | 100 W/m²K (cooling by forced air) |
| $\varepsilon$ | 0.09 |
| $\sigma$ | $5.67 \times 10^{-8}$ W/m²K⁴ |

Eqn. 1 is discretized using implicit finite difference scheme and solved using TDMA[14]. The radial control volume size and time step are Δr=3.5mm and Δt=5s. The values of the various parameters used in the study are listed in Table 1. The heat transfer coefficients are the estimated values for air with free convection and air with forced convection during heating and cooling the billet, respectively[15] which is the case during industrial processing conditions. The microstructures at *r*=0, *r*=R/2 and *r*=R after (i) casting, (ii) homogenization, and (iii) post-homogenization cooling are compared in this study.

The temperature profile at the three positions during the proposed homogenization schedule for AA7050[11] are compared in Figure 3. The temperatures at the 3 positions (Figure 2) during heating, holding, and cooling does not vary much for a billet of radius 0.35m. The Biot number, $Bi(=hr/k_c)$ which indicates the dominant heat transfer mode, is 0.023 and 0.23 during heating and cooling respectively. $Bi \ll 1$ indicates heat conduction in the billet offers little resistance to heat transfer and the temperature difference in the body is small compared to the external temperature difference as seen in the results. However, geometry would have made a difference for a larger sized billet.

3. Results and Discussion
3.1. Radial Variation in Microstructure
3.1.1. Initial As-cast Microstructure

The initial as-cast microstructure was based on the predictions by the DC-Cast solidification model for a AA7050 alloy of nominal composition Al-6.2Zn-2.3Cu-2.25Mg-0.115Zr. The mixture composition and local solidification time (LST) were taken at three different radial locations at an axial height of 1.5m from a billet of height 3m. Figure 4 shows the mixture composition for Zn, Cu, and Mg, the LST, and the calculated secondary dendrite arm spacing for the surface, mid-radius, and centerline. The LST was calculated based on when the control volume started and finished solidification and does not take into consideration the movement of solid particles. To take account of solid motion, the LST is assumed to be within 10% of the value predicted by the model. The relationship between LST, $t_f$ and SDAS, $\lambda_2$ was taken from Dantzig and Rappaz[16], which is valid for Al alloys where 0.1 s < LST < 107 s.

$$\lambda_2 = K t_f^{1/3} \qquad (4)$$

where, $K = \frac{10^{-5} m}{s^{\frac{1}{3}} for}$. For convection-controlled growth the constant in the above expression is closer to $10^{-9}$ and the exponent for $t_f$ is closer to ½.

Figure 4 shows the predicted LST and macrosegregation of Zn in a DC-cast AA7050 alloy as predicted by the solidification model. Figure 5(a) shows the composition of the billet at the 3 positions studied: centerline, mid-radius, and surface of the billet. Positive macrosegregation is at mid-radius and surface of the billet. Not much macrosegregation is observed for Zr which is present in trace amounts. Figure 5(b) shows the predicted LST and calculated SDAS at the three positions. The solidification time at the surface of the billet is low compared to that at the centerline due to heat transfer which is fast at the surface leading to smaller SDAS.

The predicted compositions and SDAS lengths are used to estimate the as-cast microstructures in the meso-scale for input to the homogenization model. Figure 6(a) shows the as-cast secondary phase fractions as predicted by Thermo-Calc$^{TM}$ corresponding to compositions shown in Figure 5(a). The precipitation of primary $Al_3Zr$, S ($Al_2CuMg$), V(solution of $Mg_2Zn_{11}$ and $Al_5Cu_6Mg_2$) and T($Al_2Mg_3Zn_3$) phases are predicted by Thermo-Calc$^{TM}$. As the compositions and temperatures of the mid-radius and surface positions are the same, the volume fraction of the secondary phases is also the same. The secondary phases are more at the surface which has a higher composition. However, the primary $Al_3Zr$ is higher at the centerline due to higher Zr composition at the centerline which follows a macrosegregartion pattern reverse of other elements due to its partition coefficient >1 in Al.

The predicted compositions are also used in Figure 6(b) and (c) to show the microsegregation predicted by Thermo-Calc$^{TM}$ at the centerline and surface of the billet respectively. The amount of Zn, Cu and Mg is higher at the mid-radius and surface positions, leading to higher amounts of these elements across the grain, while Zr is higher at the centerline.

The initial microstructure for the homogenization model at each position is represented by the 1D half-grain domain described by Priya et al.[10] with the predicted microsegregation of elements across the half-grain and interdendritic phases in the interdendritic 1$^{st}$ cell. The T and V are taken as a single phase and they both transform to the S phase.

### 3.1.2. Homogenized Microstructure

Microstructure changes at the grain boundaries and across the grain during homogenization. The transformation and dissolution of the T, V and S phases occurs at the grain boundaries while nano-sized coherent metastable $Al_3Zr$ are precipitated across the grain. Due to different compositions across the radius of the billet which leads to a difference in the as-cast interdendritic phase fractions, there is a variation in the transformation and dissolution kinetics as shown in Figure 7. The three homogenization steps in the proposed homogenization schedule[10] are 10 hours at 420°C to precipitate dispersoids followed by a second step at 470°C to dissolve T+V and a third at 480°C to minimize S is provided.

During step I, the T+V transform to S. After step I, the amount of T+V and S at the centerline is less than that at the mid-radius and surface positions because of the lower as-cast volume fraction of T+V. During step II, the T+V completely dissolve at the centerline in 3.5 hrs while they dissolve at the surface and mid-radius positions in 4 hrs as seen in Figure 7(a). If we move on to the next step without allowing for complete dissolution of T+V across the entire cross section of the billet, these phases might melt as they have a low melting temperature. To ensure the T+V phases do not melt, we need to move on to the next step after 4 hours. Redissolution of some of the precipitated S phase is observed ('reversion') during heating from 420°C to 470°C and from 470°C to 480°C as seen in Figure 7(b). Step III involves dissolution of the S phase in which the billet must be heated for more than 10 hours.

Dispersoids of $Al_3Zr$ precipitate in the grains across the cross section of the billet. The centerline is Zn, Cu and Mg lean and Zr rich compared to the mid radius and the surface. Solute lean and Zr rich centerline has more primary $Al_3Zr$ as seen in Figure 6(a) still leaving Zr during microsegregation comparable to the surface as seen in Figure 6(c). As seen in Figure 7(c), the number densities and mean radii of the dispersoids are both higher at the centerline than at the surface and mid-radius positions due to more nucleation and higher growth rates due to higher supersaturation of Zr at the centerline indicating the influence of other solute atoms in the alloy on the solid solubility of Zr. This leads to higher probability of recrystallization at the periphery than at the center of the billet. This is also observed in industrial extrusions, where peripheral coarse grained microstructures[17] is a major problem.

### 3.1.3. Post-homogenization Cooled Microstructure

Industrial practice of cooling the billet involves forcing air over its surface after the furnace is switched off. As seen in Figure 3, this practice causes cooling at a rate of approximately 148°C/hr at all the three radial positions. The precipitation model is used to simulate the microstructural evolution during post-homogenization cooling. The precipitation of S ($Al_2CuMg$), η ($MgZn_2$), T ($Al_2Mg_3Zn_3$) and Θ($Al_2Cu$), in the decreasing order of temperature, is predicted by the model at all the radial positions.

The number density of the η and T phases are more towards the surface than at the center, due to higher Zn at the mid-radius and surface positions, while the undesirable S phase is more at the centerline, as seen in Figure 8(a). The mean lengths of these platelets are longer and volume fractions are higher at the surface as seen in Figure 8(b) and (c) due to higher solute available for growth. As seen in Figure 8(d) the size of the S platelets at the surface exceeds 0.6μm and may not dissolve during pre-heat and may cause melting at the surface which may affect the surface finish[18]. Also larger precipitates at the surface may lead to particle stimulated nucleation of recrystallized grains[19] leading to inhomogeneous mechanical properties across the cross section of the billet. The solute remaining in the matrix is higher at the surface leading to more extrusion pressure required for extrusion[20]. The precipitation during cooling, in general produces inhomogeneous precipitation leading to inhomogeneous mechanical properties across the cross section of the billet which is undesirable.

### 3.2. Process Recommendations

Based on the study, a billet of radius 0.35m needs to be homogenized for 10 hrs at 420 °C to precipitate dispersoids, 4 hrs at 470°C to dissolve T+V and more than 10 hrs at 480°C to minimize S phase. Sampling at the mid-radius and surface is necessary to determine a proper homogenization schedule as they have a higher composition leading to higher volume fractions of secondary interdendritic phases which take longer to dissolve during homogenization step II of dissolving the T+V phases. Increasing the temperature to 480°C before T dissolves in the entire cross-section may cause melting at the surface and mid-radius causing pores degrading the mechanical properties of the billet.

The Zr macrosegregation in the billet causes higher Zr at the centerline position leading to higher $Al_3Zr$ dispersoid number density compared to the surface making the surface more prone to recrystallization. Addition of a trace element with a macrosegregation profile reverse of that of Zr would solve the problem both at the macro and the micro scale. Addition of Scandium, which has a partition coefficient less than 1 (whereas $k_{Zr}>1$) is a viable solution to the problem leading to macrosegregation and microsegregation patterns reverse of Zr[21]. A $k_{Sc} <1$ would lead to segregation reverse of Zr and precipitate $L_{12}$ $Al_3Sc$ and $Al_3(Sc,Zr)$ in regions lean in Zr. The $Al_3(Sc,Zr)$ dispersoids are nano-sized and coherent and more efficient than $Al_3Zr$ in increasing the strength.[21] Thus, addition of Sc may not only lead to uniform mechanical properties across the cross section of the billet, but may also reduce dispersoid free zones by precipitating $Al_3(Sc,Zr)$ type precipitates close to the grain boundary. Sc in the range of 0.18-0.2% can be added to Zr (0.1-0.2%) containing 7XXX alloys.

Precipitation during cooling under industrial conditions, produces precipitates more in number and larger in size (> 0.6 μm) at the surface than at the centerline position. The larger precipitates may induce particle stimulated nucleation of the recrystallization which can be inhibited by having more dispersoids at these positions. This is possible by having Sc in the alloy as described above. The particles may also not dissolve affecting the surface finish for which a cooling rate higher than the general industrial practice is needed.

## 4. Conclusion

The radial variation of microstructure during DC-casting and homogenization of a cylindrical billet of radius 0.35m has been studied. The DC casting model by Fezi et al.[9] and homogenization model from Priya et al. [10,11] have been used to characterize the radial variation of microstructure. Macrosegregation causes difference in compositions across the cross section of the billet leading to higher compositions and interdendritic phases at the mid-radius and surface positions. 10 hrs at 420°C, 4 hrs at 470°C and more than 10hrs at 480°C leads to homogenization across the entire billet without remelting any of the interdendritic phases when taken to higher temperatures. This schedule matches the scheduled proposed for this alloy in reference [10]. The lower Zr content at the surface leading to lower number densities of $Al_3Zr$ dispersoids, so these regions are more prone to recrystallization. Addition of Sc in the range of 0.18-0.2%, might lead to more uniform microstructure and mechanical properties across the grains and across the cross section of the billet.  Post-homogenization cooling under industrial conditions leads to larger precipitates at the surface which may cause particle stimulated nucleation of recrystallized grains or may even remain undissolved during preheat causing melting. Using higher cooling rates can reduce the size of the precipitates. However, higher cooling rates are only possible for smaller sections.

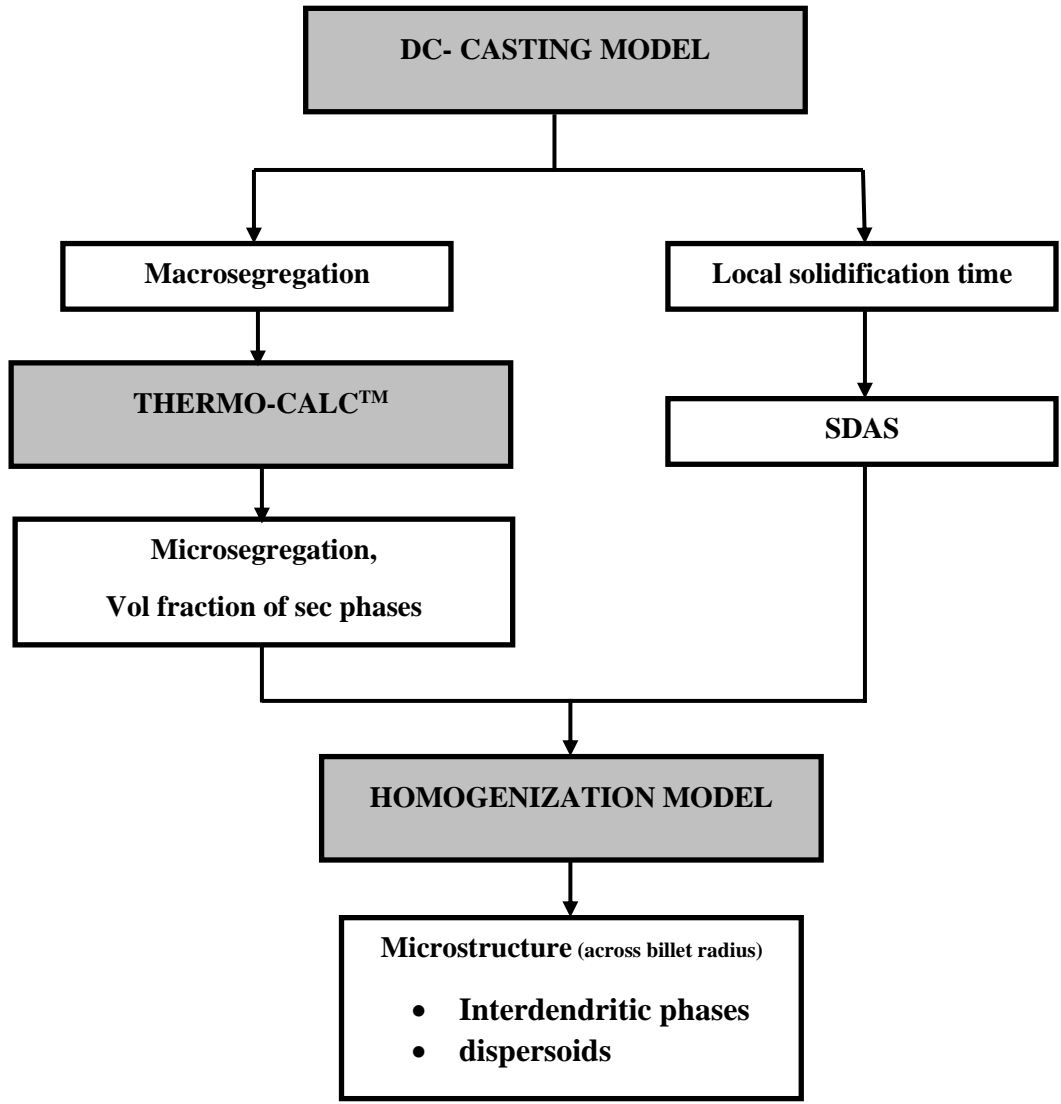

**Figure 1**: Schematic showing the inputs and outputs of the numerical models involved to study the microstructural evolution during homogenization and cooling. The flow of the simulations is illustrated in Figure 1.

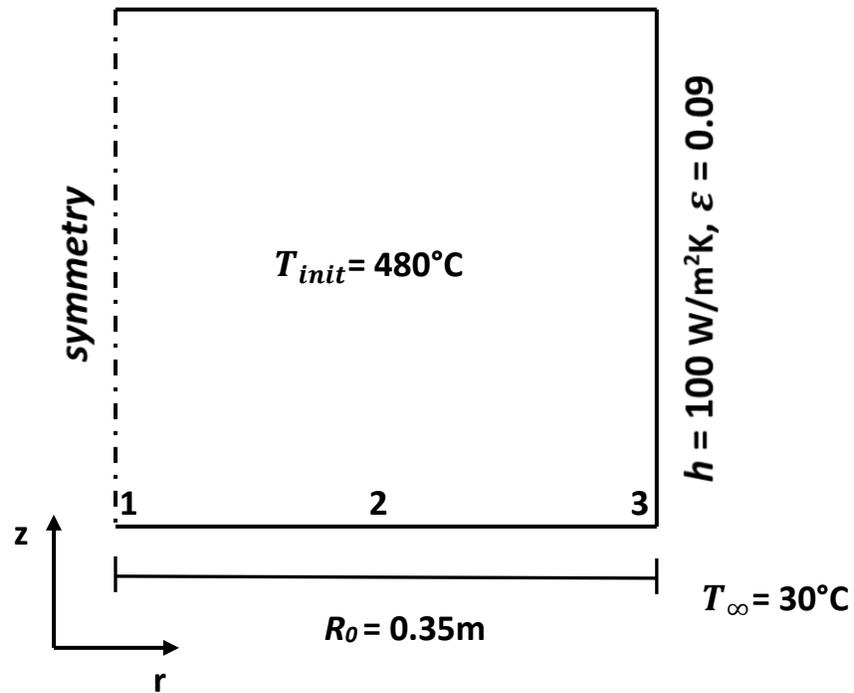

**Figure 2**: Schematic showing the axisymmetric domain and the boundary conditions during industrial cooling of the billet. The positions where the microstructures are compared are numbered. The initial temperature is ambient temperature at the beginning of the heating cycle.

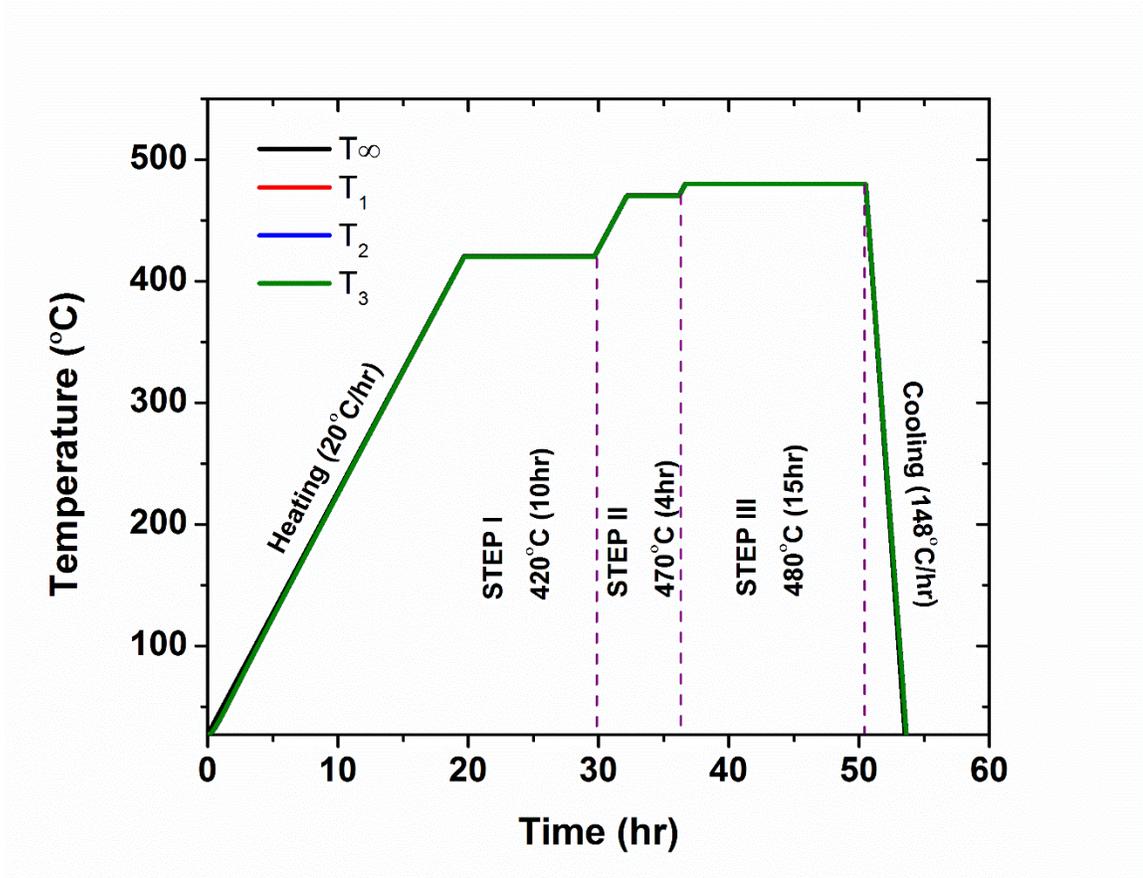

**Figure 3**: The homogenization heating, holding, and cooling cycle chosen in the study showing little variation in temperatures at the 3 positions studied.

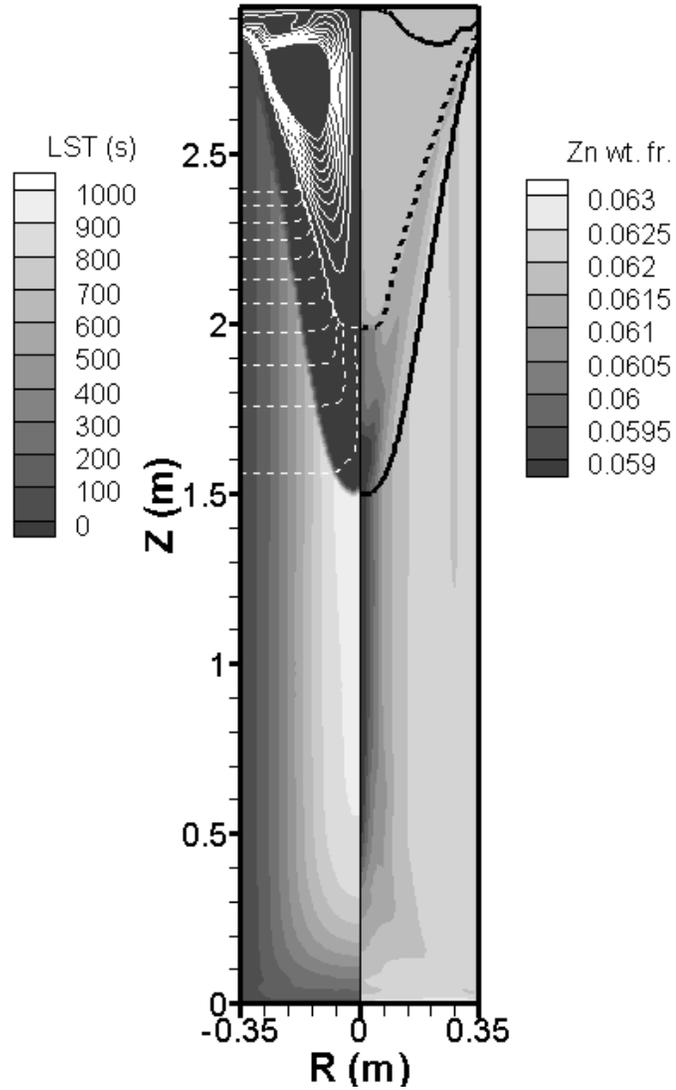

**Figure 4**: Contour plot of the local solidification time (LST) in seconds and the Zn composition fields for a steady state casting velocity of 90 mm/min. the Zn composition and sump are on the right, with solid lines representing the liquidus and solidus and the dotted line packing location of the free-floating solid. On the left are liquid flow solid streamlines with $0.1 < \rho\Psi < 1$ and $\Delta\rho\Psi=0.1$ kg/s, dotted streamlines with $-0.005 < \rho\Psi < -0.000025$ and $\Delta\rho\Psi=0.0004975$ kg/s and LST.

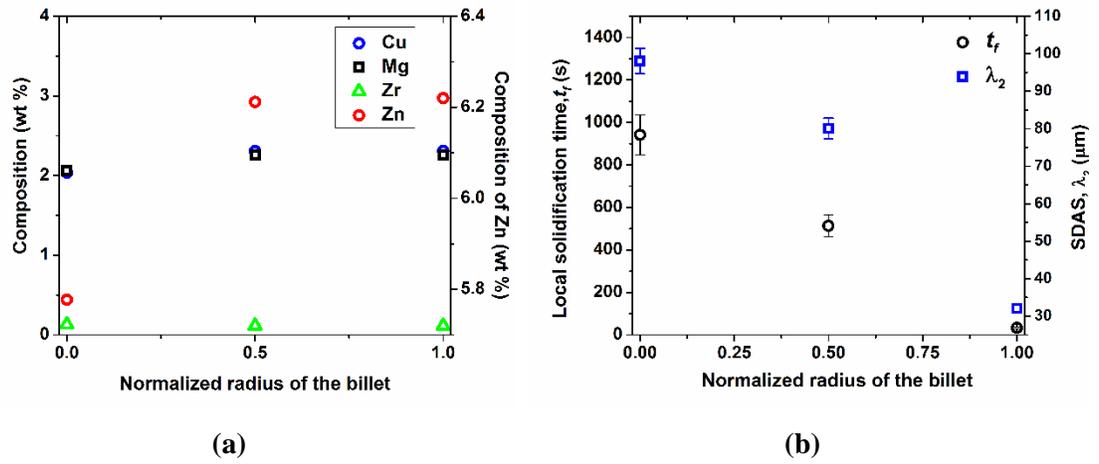

**Figure 5**: The predicted (a) compositions and (b) LST and SDAS across the radius of the billet.

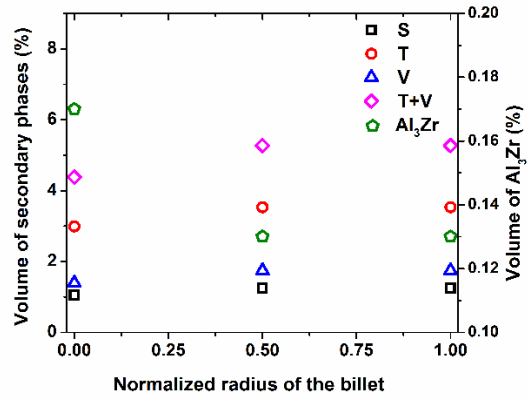

**(a)**

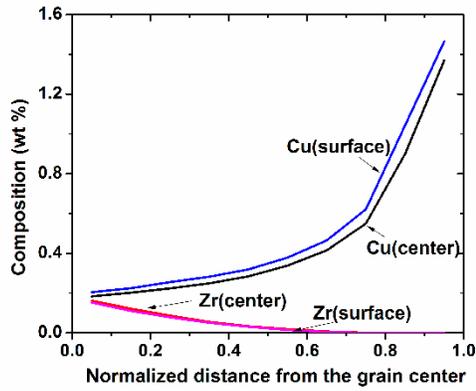

**(b)**

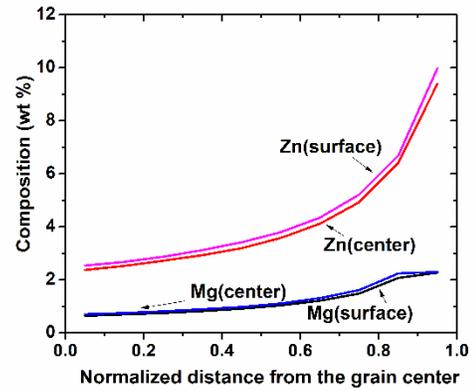

**(c)**

**Figure 6**: (a) Predicted as-cast volume fractions of secondary phases at different radial positions in the billet; The microsegregation across the grains for (b) Cu and Zr and (c) Zn and Mg.

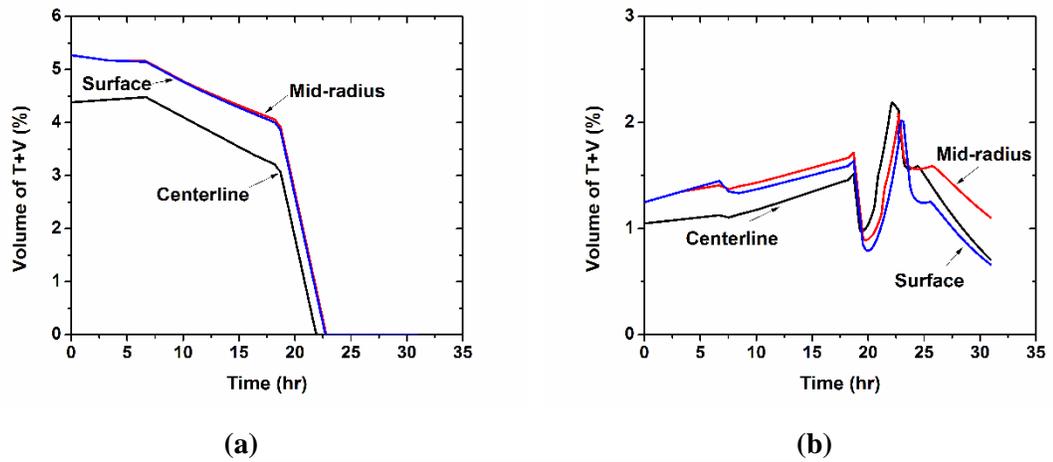
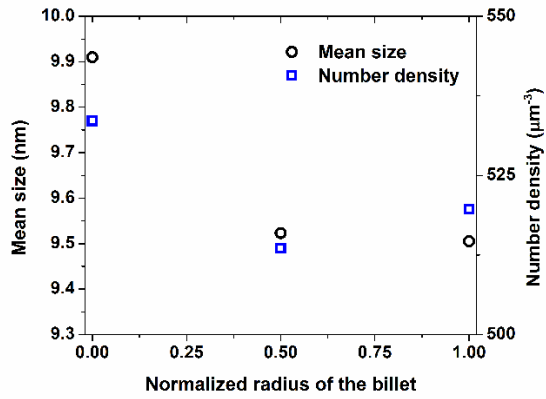

**Figure 7**: (a) Evolution of the T+V phases; (b) S phase during homogenization; (c) The number density and mean radii of the dispersoids across the radii of the billet after Step II of homogenization.

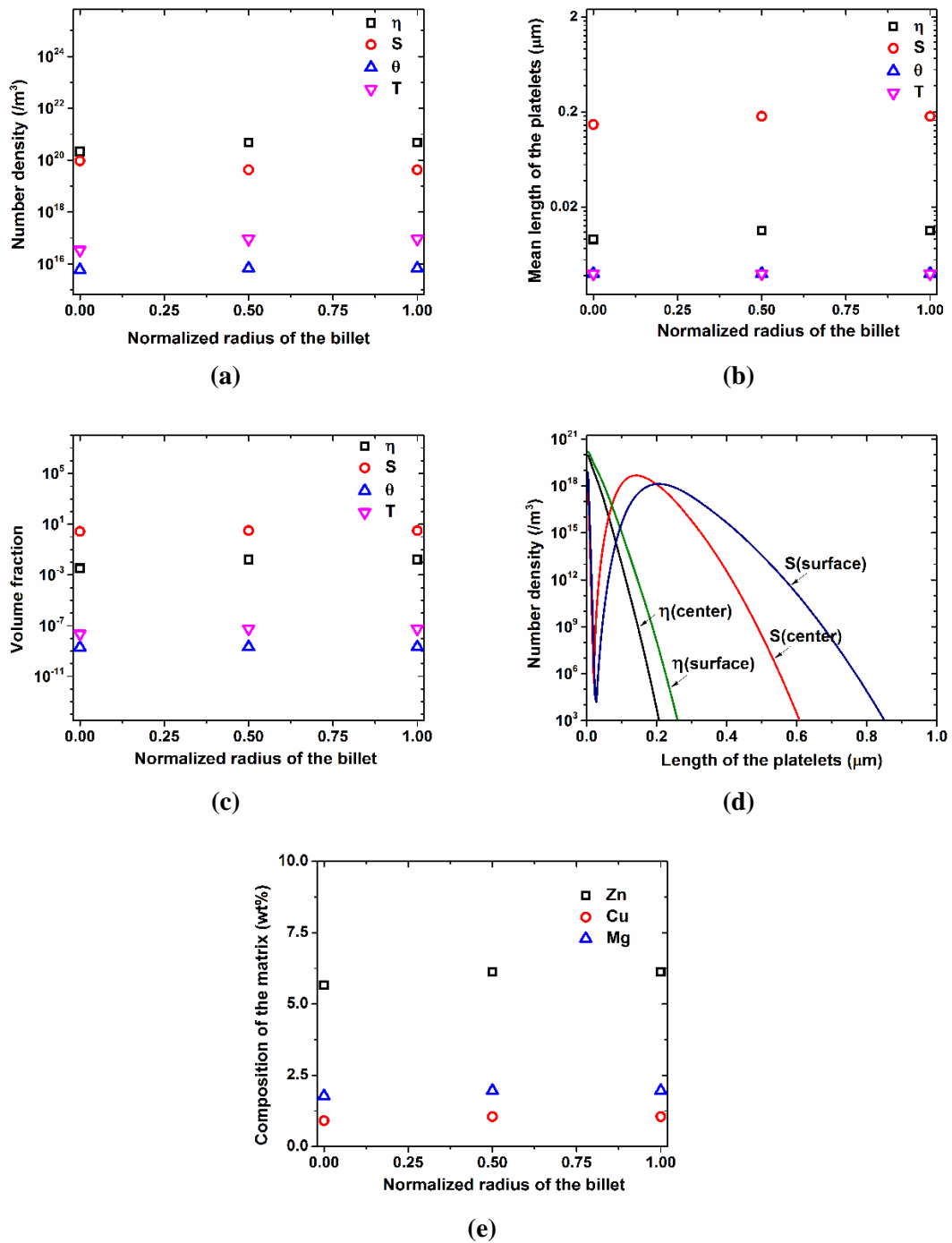

**Figure 8**: The radial variation of (a) number density, (b) mean length of the platelets, (c) volume fraction, (d) size distribution of the phases precipitated during cooling under industrial conditions and (e) radial variation of the composition of the matrix.